\documentclass[preprint2]{aastex}
\usepackage{float}

\shorttitle{Extinction from [FeII] near-infrared emission lines}
\shortauthors{Pecchioli et al.}

\begin{document}

%% LaTeX will automatically break titles if they run longer than
%% one line. However, you may use \\ to force a line break if
%% you desire.

\title{Deriving the extinction to young stellar objects using [FeII] near-infrared emission lines. \\
    Prescriptions from GIANO high-resolution spectra}

%% Use \author, \affil, and the \and command to format
%% author and affiliation information.
%% Note that \email has replaced the old \authoremail command
%% from AASTeX v4.0. You can use \email to mark an email address
%% anywhere in the paper, not just in the front matter.
%% As in the title, use \\ to force line breaks.

\author{T. Pecchioli}
\affil{Dipartimento di Fisica e Astronomia, Universit\` a degli Studi di Firenze,
    via G. Sansone 1, 50019 Sesto Fiorentino (Firenze), Italy}

\author{N. Sanna, F. Massi and E. Oliva}
\affil{INAF - Osservatorio Astrofisico di Arcetri, Largo E. Fermi 5,
              50125 Firenze, Italy}
             \email{sanna,fmassi,oliva@arcetri.astro.it}

%% Notice that each of these authors has alternate affiliations, which
%% are identified by the \altaffilmark after each name.  Specify alternate
%% affiliation information with \altaffiltext, with one command per each
%% affiliation.

%% Mark off your abstract in the ``abstract'' environment. In the manuscript
%% style, abstract will output a Received/Accepted line after the
%% title and affiliation information. No date will appear since the author
%% does not have this information. The dates will be filled in by the
%% editorial office after submission.

\begin{abstract}
The near-infrared emission lines of Fe$^{+}$ at $1.257$, $1.321$, and $1.644$ $\mu$m
share the same upper level; their ratios can then be exploited to derive the extinction
to a line emitting region once the relevant spontaneous emission coefficients are known.
This is commonly done, normally from low-resolution spectra, 
in observations of shocked gas from jets driven by Young Stellar Objects.  
In this paper we review this method,
provide the relevant equations,
 and test it by analyzing high-resolution ($R \sim 50000$) 
near-infrared spectra of 
two young stars,
namely the Herbig Be star HD~200775 and the Be star V1478~Cyg,
which exhibit intense emission lines. The spectra were obtained with the new GIANO 
echelle spectrograph at the Telescopio Nazionale Galileo.
Notably, the high-resolution spectra allowed checking the effects of overlapping 
telluric absorption lines. 
A set of various determinations of the Einstein coefficients are compared to show how much
the available computations affect extinction derivation. The most recently
obtained values
are probably good enough to allow reddening determination within 1 visual mag
of accuracy. 
Furthermore, we show that [FeII] line ratios from low-resolution 
pure emission-line spectra
in general are likely to be in error due to the impossibility to properly account for 
telluric absorption lines. 
If low-resolution spectra are used for reddening determinations, 
we advice that the ratio $1.644/1.257$, 
rather than $1.644/1.321$, should be used, 
being less affected by the effects of telluric absorption lines.
\end{abstract}

%% Keywords should appear after the \end{abstract} command. The uncommented
%% example has been keyed in ApJ style. See the instructions to authors
%% for the journal to which you are submitting your paper to determine
%% what keyword punctuation is appropriate.


%% From the front matter, we move on to the body of the paper.
%% In the first two sections, notice the use of the natbib \citep
%% and \citet commands to identify citations.  The citations are
%% tied to the reference list via symbolic KEYs. The KEY corresponds
%% to the KEY in the \bibitem in the reference list below. We have
%% chosen the first three characters of the first author's name plus
%% the last two numeral of the year of publication as our KEY for
%% each reference.

%% Authors who wish to have the most important objects in their paper
%% linked in the electronic edition to a data center may do so by tagging
%% their objects with \objectname{} or \object{}.  Each macro takes the
%% object name as its required argument. The optional, square-bracket 
%% argument should be used in cases where the data center identification
%% differs from what is to be printed in the paper.  The text appearing 
%% in curly braces is what will appear in print in the published paper. 
%% If the object name is recognized by the data centers, it will be linked
%% in the electronic edition to the object data available at the data centers  
%%
%% Note that for sources with brackets in their names, e.g. [WEG2004] 14h-090,
%% the brackets must be escaped with backslashes when used in the first
%% square-bracket argument, for instance, \object[\[WEG2004\] 14h-090]{90}).
%%  Otherwise, LaTeX will issue an error. 

\section{Introduction}

Emission lines are instrumental in deriving the physical conditions of
nebular sources associated with young stars
in the Galaxy. E. g., \citet{nisini05} showed that
the physical conditions in Herbig-Haro objects and jets from young stellar
objects (YSOs) can be obtained from various diagnostic line ratios
emitted by a number of atoms and ions in a wavelength interval
ranging from the optical to the near-infrared (NIR).
Out of these, the [FeII] line are of particular interest. Jets from
YSOs are usually very rich in [FeII] lines at
ultraviolet, optical, and NIR wavelengths, allowing sampling of regions
in a wide range of physical conditions (Giannini et al. 2013,
Giannini et al.\  2015b).
Hollenbach \& McKee (1989)
already predicted that [FeII] lines at $1.7$ and $1.3$ $\mu$m are
good tracers of $J$ shocks in dense ($\gtrsim  10^{5}$ cm$^{-3}$) gas.

However, jets from YSOs are usually heavily embedded in gas and dust,
thus an accurate knowledge of the extinction is needed to fully exploit
the diagnostic power of multiple line ratios. In this respect, the
[FeII] lines at $1.6440$, $1.3209$, and $1.2570$ $\mu$m all arise from
the same upper level, i. e. $3d^{6}(^{5}D)4s$ ($a^{4}D_{7/2}$),
and in principle can be used to derive extinction 
provided the spontaneous emission coefficients are known with good
accuracy. \citet{nisini05} noted that the ratio [FeII] $1.644/1.257$
yielded a systematically
higher (by roughly a factor of 2) extinction
compared to the $1.644/1.321$ ratio in their low resolution spectra of
jet HH1. In addition, they checked that the $1.644/1.257$
ratio led to $A_{V}$ values higher than estimated from optical
observations for a number of HH objects. They
suggested that these facts point to inaccuracies in the theoretical
transition probabilities available. Massi et al. (2008) cautioned against the
possibility that the discrepancy between ratios could arise from inaccuracies in the
telluric absorption correction caused by the $1.257$ $\mu$m line
falling in the region of the $^{1}\Delta_{g}$ absorption band of
atmospheric O$_{2}$. Giannini et al.\ (2015a) tried determining the
Einstein spontaneous emission rates of the three Fe$^{+}$ lines using X-shooter
spectra of HH1, still finding a few cases of discrepancies between their determination
and data from the literature.

During one of the commissioning cycles with the NIR high-resolution
spectrometer GIANO, which has been available at the Telescopio Nazionale Galileo
since 2013, we obtained spectra of a small sample of bright Be and Herbig Ae/Be
stars as a test case. These are relatively extincted objects, so we are
interested in deriving their reddening with high accuracy. Thus, we
have decided to further investigate the issue of the [FeII] lines by both exploiting 
the high-spectral resolution capabilities provided by GIANO,
and considering the most recent computations of the Einstein coefficients
(Bautista et al.\ 2015). We have also examined the effects of using different
extinction laws. The fact that the emission lines are superimposed on continuum
spectra allows one to also have a simultaneous view of the telluric
absorption lines.
Our aim was ultimately to obtain a few prescriptions on extinction and Einstein coefficient
determination from NIR spectroscopic observations of the [FeII] lines.
In this paper we review this method using the high-resolution spectra as templates, discussing 
the relevant formulae involved and the expected error budgets. 
The paper layout is the following. In Sect.~2 we outline the observations
and data reduction. We discuss the issues related with flux calibration in Sect.~3 
and correction of the line integrated fluxes for telluric line absorption in Sect.~4.
In Sect.~5 we explain how
to derive extinction from line ratios and discuss the effects of systematic errors. 
In Sect.~6 we analyze the implications pointed at by our data. Finally, Sect.~7 is a summary of 
our results.
Some useful additional documentation is in appendix. Appendix~A gives some more details on
data reduction, Appendix~B discusses some flux calibration issues, and 
Appendix~C contains three figures showing the typical telluric absorption
spectra in wide spectral windows around the [FeII] line wavelengths.

\section{Observations and data reduction}

%% In a manner similar to \objectname authors can provide links to dataset
%% hosted at participating data centers via the \dataset{} command.  The
%% second curly bracket argument is printed in the text while the first
%% parentheses argument serves as the valid data set identifier.  Large
%% lists of data set are best provided in a table (see Table 3 for an example).
%% Valid data set identifiers should be obtained from the data center that
%% is currently hosting the data.
%%
%% Note that AASTeX interprets everything between the curly braces in the 
%% macro as regular text, so any special characters, e.g. "#" or "_," must be 
%% preceded by a backslash. Otherwise, you will get a LaTeX error when you 
%% compile your manuscript.  Special characters do not 
%% need to be escaped in the optional, square-bracket argument.

\begin{figure}
   \centering
   \includegraphics[width=7cm]{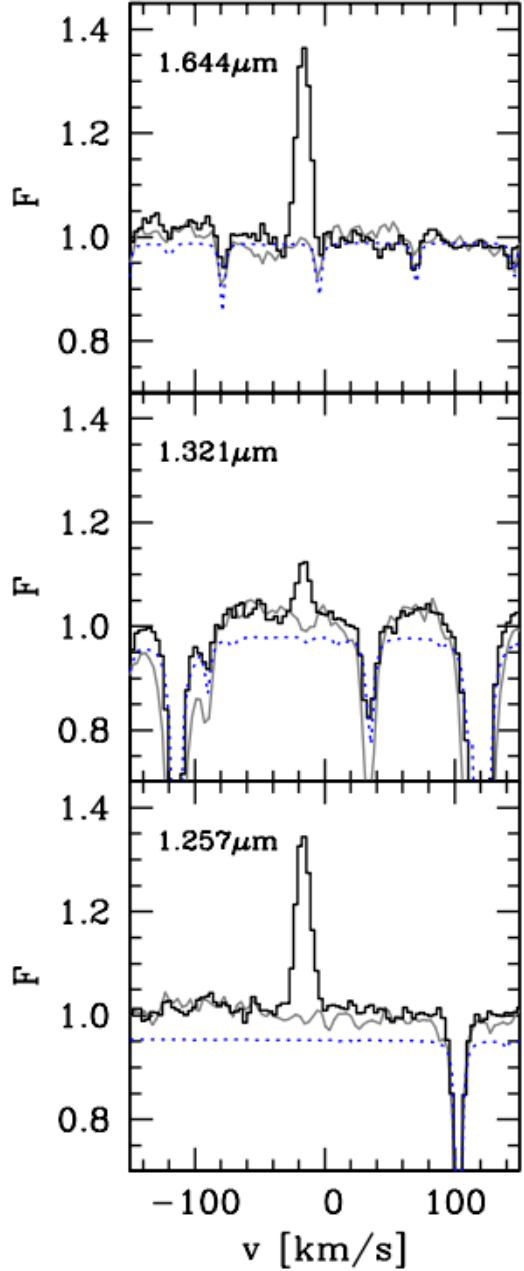}
   \caption{[FeII] emission lines (solid histogram) towards HD~200775
at $1.644$ $\mu$m (upper panel),
$1.321$ $\mu$m (middle panel) and $1.257$ $\mu$m (bottom panel).
The solid gray line shows the spectrum of the telluric star Hip~89584 around the [FeII] lines,
while
the dotted blue line outlines the telluric transmission
obtained by an atmospheric model with a spectral resolution R=100000.
All fluxes are plotted as a function of radial velocity $V$, assuming $V = 0$ for
a line emitted at the systemic velocity, and are normalized to the continuum.
              \label{fig_HD200775_FeII}}
    \end{figure}

\begin{figure}
   \centering
   \includegraphics[width=7cm]{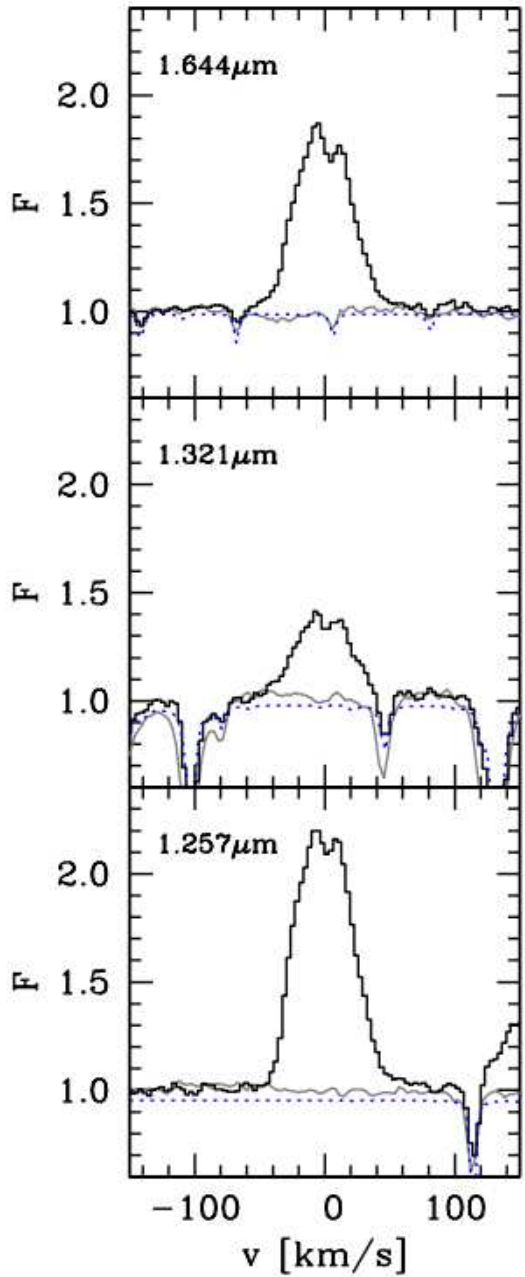}
   \caption{Same as Fig.~\ref{fig_HD200775_FeII}, but for V1478~Cyg.}
              \label{fig_V1478Cyg_FeII}%
    \end{figure}

The stars that we use as benchmarks are three bright objects 
classified as Herbig Ae/Be star candidates in the literature,
namely \object{HD~200775}, \object{V1478~Cyg}, and \object{V1686~Cyg}.
These stars are characterized by strong
emission lines originated in the circumstellar environment. Their
spectra were obtained with the high-resolution NIR spectrometer GIANO
(R$\approx$50000) in the night of July 30, 2013, during the instrument commissioning phase.
GIANO is a cross-dispersed spectrograph covering the wavelength range between
$0.97-2.4$ $\mu$m in a single exposure, via 49 spectral orders. Except a few narrow
spectral intervals at wavelengths $> 1.7$ $\mu$m, most of the band
is imaged on a $2048 \times 2048$ pixels, HgCdTe detector.
Two ZBLAN fibers with a sky-projected diameter of
$\sim 1$ arcsec (hereby A and B) are used to transfer two spots $\sim 3$ arcsec apart
from the telescope focal plane to the slit, so that two spectra are obtained
simultaneously. More details can be found in \citet{origlia13}.
We took spectra of the stars by the nodding-on-fiber technique, i.e. by observing the object
alternatively through fiber A and fiber B with the same exposure time.
This technique allows canceling the background emission and any detector biases by
simply subtracting frame B from frame A. The integration time of each spectrum was of 300~s.
Further details on data reduction are given in Appendix~A.
Unfortunately, the $1.321$ $\mu$m line is not detected (or only barely detected
above the noise) and the $1.644$ $\mu$m is clearly
affected by a telluric absorption line in the spectrum of V1686~Cyg. 
As a consequence, we will use the data from this
star only to assess the goodness of the intercalibration in Appendix~B, 
being the [FeII] line correction extremely critical.
The spectra extracted in the wavelength ranges
of interest, normalized to the continuum, are shown in Figs.~\ref{fig_HD200775_FeII} and
\ref{fig_V1478Cyg_FeII}.

%
%
% BEGINNING FIGURES

%     \begin{figure}
%   \centering
%   \includegraphics[width=7cm]{righe_V1686Cyg.ps}
%   \caption{Same as Fig.~\ref{fig_HD200775_FeII}, but for V1686~Cyg.}
%              \label{fig_V1686Cyg_FeII}%
%    \end{figure}
%

%% In this section, we use  the \subsection command to set off
%% a subsection.  \footnote is used to insert a footnote to the text.

%% Observe the use of the LaTeX \label
%% command after the \subsection to give a symbolic KEY to the
%% subsection for cross-referencing in a \ref command.
%% You can use LaTeX's \ref and \label commands to keep track of
%% cross-references to sections, equations, tables, and figures.
%% That way, if you change the order of any elements, LaTeX will
%% automatically renumber them.

%% This section also includes several of the displayed math environments
%% mentioned in the Author Guide.

\section{
Intercalibration of line fluxes
}
\label{intercal}

Accurate line ratio derivation needs accurate flux or line ratio calibration.
In principle
this can be obtained either from the source spectra themselves, if known, or 
by observing stars during the night whose NIR spectra are known with
accuracy (spectrophotometric standards). 
%If the flux is calibrated using the target continuum spectrum itself, source 
%variability can cause a systematic error on the line ratios.
%To assess the effects of this calibration error, one has to take into account
%that $F_{\lambda}$ in Eq.~\ref{next:eq} is actually $F_{{\rm cal},\lambda}$,
Basically, if the underlying target continuum is used, the calibrated flux is given by:
\begin{equation} 
\label{cal:flux:line}
F_{{\rm cal},\lambda} = F_{{\rm meas},\lambda} \times \frac{A\lambda^{\alpha}}
{C_{\lambda}}
\end{equation}
where $F_{{\rm meas},\lambda}$ is the measured line integrated flux (in counts times wavelength),
$C_{\lambda}$ is the measured continuum flux (in counts) outside the line, and $A\lambda^{\alpha}$ is the source continuum flux density (i. e.,
${\rm d}F/{\rm d}\lambda$) from a fit, e. g.,  to the 2MASS photometry. 

One problem of using the target continuum is a possible spectral variability of our sources,
in addition there are no observations of spectrophotometric standards available
during the night, either.
However, we note that, as far as line {\em ratios} are
requested, what is actually needed is an intercalibration between different wavelength ranges
rather than flux calibration,
which is less demanding. This can be achieved by using spectra of telluric standard stars
and/or stars of known spectra which do not exhibit variability. 
In practice, 
the measured line ratio $F_{\rm meas}(\lambda_1)/F_{\rm meas}(\lambda_2)$,
where $F_{\rm meas}(\lambda)$ is the {\it integrated} line flux (in counts by
wavelength),
can be calibrated multiplying it by a line-ratio calibration factor given by:
\begin{equation}
\label{cal:ratio:line}
X = \frac{F_{\rm meas,s}(\lambda_2)}{F_{\rm meas,s}(\lambda_1)} \frac{S_{\rm s}(\lambda_1)}{S_{\rm s}(\lambda_2)}
\end{equation}
where $F_{\rm meas,s}(\lambda)$ is the flux measured at $\lambda$ (in counts) from the continuum of a star of known
spectrum and $S_{\rm s}(\lambda)$ is its theoretical continuum flux density at $\lambda$.
This method has the advantage that the effects of changes of seeing and air mass between
target and calibrator integrations mostly cancel out in the ratio.

Unfortunately, only the three targets were observed 
during the night of 30. Nevertheless, a few suitable sources were observed on the nights before and
after. This should not be of major concern,
since one can expect that, as a first approximation, {\it continuum} telluric absorption
only depends on the atmospheric gas column density. Thus, 
line-ratio calibration factors should remain almost constant in time, at least on 
nearby nights. What is bound to exhibit changes is the absorption in the narrow wavelength
regions dominated by telluric absorption lines. 
Therefore, normalized flux ratios obtained from stars of known spectra
can be used as correcting factors for 
calibrating line {\it ratios} in the target spectra, as long
as the emission lines of interest do not overlap strong telluric absorption lines.

We have checked the consistency of line-ratio calibration factors in time by using
spectra of a telluric standard star, namely  \object{Hip~89584} (an O6.5 V star), of 
\object{VEGA}
(A0 V, two observations  5 hr apart),
of \object{HD~188001} (O7 Iabf), and of \object{HD~190429A} (O4 If) taken on the night of 29 July, 
plus two spectra of VEGA
(7 hrs apart) taken on the night of 31 July. These were reduced using the same procedure as that used for
the targets. As discussed in Appendix~B, we found that the calibration factors exhibit only small
variations throughout the nights of 29 and 31, supporting our view that similar values
can be applied to the spectra taken on 30. Therefore, the line ratios were calibrated using
the mean factors derived from Hip~89584, HD~188001, and HD~190429A, listed in 
Table~\ref{ratio:cal} in Appendix~B.
The values from VEGA, although consistent with the others, were not included since its observations
were actually pointing tests.  We estimate that the calibration uncertainty on the
relevant line ratios is of order of 5-10 \%.  

%%%%%%%%%%%%%%%%%%%%%%%%%%%%%%%%%%%%%%%%%%%%%%TABELLA FLUSSI
\begin{table*}[!h]
\caption{Integrated fluxes of the [FeII] emission lines
from the stars HD~200775 and V1478~Cyg,
at $1.644$, $1.321$, and $1.257$ $\mu$m, normalized to that at $1.644$ $\mu$m. Also listed:
the line equivalent width,
full width at half maximum, and central velocity with respect to the systemic
velocity, and the extinction from the literature.} \label{flux_table}
\centering
\begin{tabular}{lcccccc}
\hline
\noalign{\smallskip}
Object & $A_{V}$ & $\lambda_c$ & Int. Flux & Eq. width & $FWHM$ & $v_r$ \\
& & ($\mu$m) & & (nm)  & (km s$^{-1}$) & (km s$^{-1}$) \\
\noalign{\smallskip}
\hline
\noalign{\smallskip}
HD~200775 & $1.8$--3$^{(a)}$ & 1.64389  & 1 & $-0.032$ & 10.76 & -16.68 \\
          & & 1.32082  & $0.32 \pm 0.03$  & $-0.006$ &  8.48 & -16.82 \\
          & & 1.25694  & $1.14 \pm 0.06$ & $-0.019$ & 11.18 & -16.42 \\
V1478~Cyg & 10--10.7$^{(b)}$ & 1.64392  & 1 & $-0.219$ & 46.81 & -1.03  \\
          & & 1.32084  & $0.26 \pm 0.01$ & $-0.074$ & 48.02 & -1.37 \\
          & & 1.25696  & $0.86 \pm 0.02$ & $-0.265$ & 48.52 & -0.50 \\
%V1686~Cyg & & 1.64393  & 1 & $-0.013$ &  4.74 & 0.60 \\
%          & & 1.32000  & $< 0.6$ & $> -0.004$ &  & \\
%          & & 1.25697  & $2.1 \pm 0.3$ & $-0.017$ & 15.48 & 1.09 \\
\hline
\end{tabular}
\tablerefs{(a) \citet{hern04};
(b) \citet{cohen85};
}
\end{table*}
%%%%%%%%%%%%%%%%%%%%%%%%%%%%%%%%%%%%%%%%%%%%%%%%%%%

\section{
Correction of absorption from telluric lines
}
\label{tell:line:corr}

Correction of absorption from telluric lines overlapping the emission lines
is as critical as a good intercalibration to obtaining accurate line ratios.
As shown in Sect.~\ref{app:cal}, a systematic error of 10 \% in the integrated
line flux is enough to cause an error of $\sim 1$ mag in the extinction. 
In this section we compare the uncertainty inherent telluric correction
in both low- and medium- or high-resolution spectra. High-resolution spectra
have the advantage of allowing an accurate mapping of the telluric lines, e. g.
through spectra of telluric standard stars. 

As far as our GIANO spectra are concerned, 
any correction for atmospheric absorption lines 
relies on the spectrum of the telluric standard Hip~89584, which was taken
on the previous night, and simulations with the ESO Cerro Paranal Advanced Sky Model
\citep[hereafter ASM]{noll}, which is based on the
HITRAN database \citep{hitran2008}.  
The spectra displayed in Figs.~\ref{fig_HD200775_FeII} and
\ref{fig_V1478Cyg_FeII}
demonstrate that all
telluric features in the Hip~89584 spectrum are retrieved by ASM, so that
we can safely assume that the location of the telluric absorption lines is
well known and can be obtained through this code in the narrow spectral windows we are studying.

Figures~\ref{fig_HD200775_FeII} and \ref{fig_V1478Cyg_FeII}
clearly show how much telluric absorption lines can affect the integrated flux measurements, as well.
The emission lines at both $1.644$ and $1.321$ $\mu$m encompass telluric absorption lines; if this
is not properly taken into account, the integrated flux will be underestimated.
%In point of fact, even a faint absorption line can have a large impact on the extinction determination.
The middle panel of Fig.~\ref{fig_HD200775_FeII},
shows that the $1.321$ $\mu$m line coincides with a faint
atmospheric absorption line. This line is very faint: measurements on the ASM
models and on the telluric star spectrum point to a change in transparency of  only 1--4
\%. 
Yet, because what is actually measured is the absorbed line {\em plus continuum} integrated flux
minus the unabsorbed continuum flux density (outside the line) by the line width,
this enhances the line flux lost. It can easily be seen that the
systematic error in the {\em measured} integrated
flux, $\Delta f_{\rm meas}$, is given by:
\begin{equation}
\frac{\Delta f_{\rm meas}}{f_{\rm meas}} \sim \frac{\Delta f}{f} + C \times
(1 - T \frac{\Delta\lambda_{\rm abs}}{\Delta\lambda_{\rm line}})
\end{equation}
where $f$ is the actual emission line integrated flux, 
$\Delta f$ the line integrated flux absorbed,
$C$ is the ratio of continuum flux density to
line peak flux density, $T$ is the minimum transparency
inside the telluric line, $\Delta\lambda_{\rm abs}$
is the width of the absorption line, and $\Delta\lambda_{\rm line}$ 
the width of the emission line. 
From Fig.~\ref{fig_HD200775_FeII} the emission line peak is less than $1/10$ of
the continuum (i.e. $C \sim 10$). Thus, the equation above shows that 
the line integrated flux may be underestimated by more than 10 \% if $\Delta\lambda_{\rm abs} \sim
\Delta\lambda_{\rm line}$.
The same problem of course affects the $1.321$ $\mu$m line towards
V1478~Cyg, but in this case the line peak is $\sim 1/2.5$ times the continuum and the
emission line is much broader, so the underestimate in integrated flux is negligible.
Note that this effect increases with increasing $C$. 
On the other hand, pure line
emission spectra ($C = 0$), like in Herbig-Haro objects, are much less affected. 

Methods to correct medium- and high-resolution NIR spectra
for telluric line absorption are described, e. g., in 
\citet{kausch}. We used a correction scheme similar to that performed
by the STSDAS routine TELLURIC\footnote{http://stsdas.stsci.edu/cgi-bin/gethelp.cgi?telluric}, 
which is based on the Beer's law, exploiting 
the telluric standard star spectrum.
This was applied to the $1.644$ and $1.321$ $\mu$m [FeII] lines and
we estimate that the effect of telluric line absorption is reduced to
a few percent of the measured integrated line flux in the $1.321$ $\mu$m
line after correction.
The [FeII] line fluxes, normalized
to the $1.644$ $\mu$m line and corrected for telluric absorption, are given in
Table~\ref{flux_table}. 

\begin{figure}
   \centering
   \includegraphics[width=14cm,angle=-90]{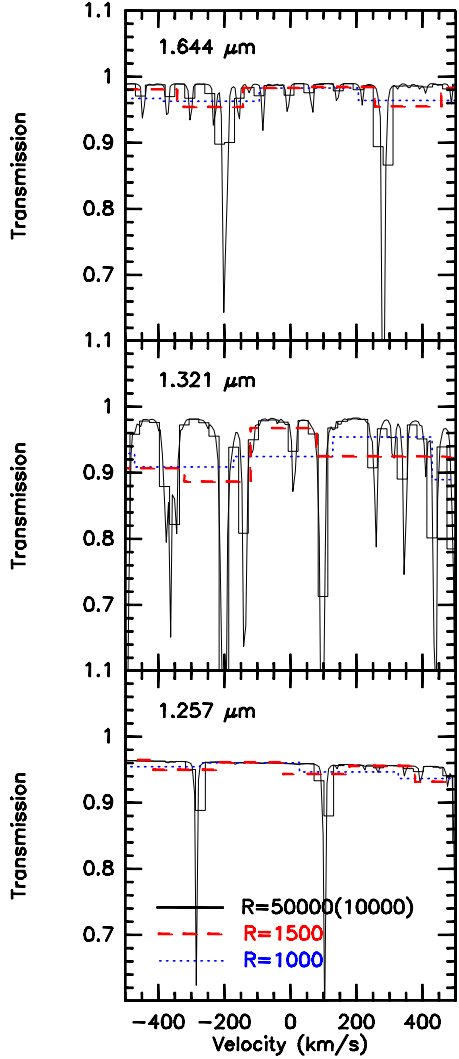}
   \caption{Atmospheric transmission model around the [FeII] emission features at
$1.644$ $\mu$m (upper panel),
$1.321$ $\mu$m (middle panel) and $1.257$ $\mu$m (bottom panel). The solid black line is
a model computed with a spectral resolution R=50000, the black histogram one computed 
with R=10000,
the dashed red line one computed with R=1500 and the blue dotted line one computed with
R=1000. The absorption lines in the upper panel are CH$_4$
and CO$_{2}$ (the faintest ones),
those in the middle panel are H$_2$O, and those in the
bottom panel are O$_2$.}
              \label{fig_plot_atmosfera}%
    \end{figure}

We point out now one possible problem with telluric absorption correction
in low-resolution spectra.
Using ASM, we have simulated the atmospheric absorption in three spectral windows
centered at the [FeII] line rest wavelengths (see Fig.~\ref{fig_plot_atmosfera}).
The wavelengths have been converted into radial velocities and the spectral windows
span an interval ranging from $-500$ to $500$ km s$^{-1}$, which should be large enough for
most galactic observations of jets from YSOs and young stars. We have overplotted a high resolution
($R = 50000$) spectrum, a medium resolution
($R = 10000$) one, and two low resolution ones ($R=1000$ and $R=1500$). The atmospheric
transmission has been estimated by assuming airmass $1.3$ and average standard values for
the remaining parameters. 
Wider spectral intervals with the high-resolution 
telluric absorption spectra as a function of wavelength are displayed in Appendix~C, Figs.~\ref{atmo_wl_1},
\ref{atmo_wl_2}, and \ref{atmo_wl_3}.
One has to keep in mind that Herbig Haro and jet feature spectra
are pure line spectra, therefore emission lines falling between telluric
absorption lines remain unaffected by them although they are widened by the
spectrograph instrumental profile at low resolution. Thus, they must be corrected using the
transmission values outside the telluric lines. But one cannot achieve this by using
continuum spectra of telluric standard stars. 
This because a correcting function is usually obtained by dividing the observed standard spectrum
by a synthetic stellar spectrum. However, the observed spectrum is not just a transmission
function of wavelength multiplied by a stellar spectrum, it is also convolved by an instrumental function.
Since the transmission function can be roughly described as a smooth continuum component plus
a number of narrow absorption lines, if the instrumental function is broad as in low resolution spectra, 
the convolution product in a point is affected by nearby absorption lines and results in a 
stronger absorption between telluric lines. 

If the target exhibits a pure emission line spectrum with no or faint underlying continuum, and the
emission line of interest lies outside any deep telluric line, the observed line profile is just the product of
the intrinsic line profile and the smooth continuum absorption component convolved by the 
instrumental function. Absorption lines are of no consequence at all in this case.
They do not enter in any way and the effective smooth absorption is not enhanced as
is the case of a star spectrum.  The effect can be roughly evaluated by looking at
the middle panel of
Fig.~\ref{fig_plot_atmosfera}; the smoothing of the telluric absorption lines due
to the low spectral resolution produces a pseudo-continuum with a lower transmission, $T'(\lambda)$, than
the actual one, $T(\lambda)$, in the regions between absorption lines. So, if the $1.321$ $\mu$m line falls between
telluric absorption lines, it will be systematically overcorrected (by a factor $T/T'$) by using the telluric
standard star technique. From Fig.~\ref{fig_plot_atmosfera}, the overcorrection can be
estimated to be $> 10$ \%, a value leading to a systematic underestimate
of $A_{V}$ $> 1$ mag, as shown by Eq.~\ref{err:eq}. But the situation is even worse,
since we have identified the telluric lines as water vapor lines, hence exhibiting the
most extreme variability. Towards some objects, the  $1.321$ $\mu$m line may coincide
with one of the telluric absorption lines, so the reverse would happen: the line flux would be
undercorrected and the extinction systematically overestimated.
Interestingly, Fig.~\ref{fig_plot_atmosfera} shows that medium resolution spectra
are unaffected by the problem and in any case telluric absorption lines can easily be
retrieved.  

On the other hand, the $1.257$ $\mu$m line is less likely to have similar problems. Only
if it coincides with one of the two O$_{2}$ lines
of the $^{1}\Delta_{g}$ band (which is centered at $1.27$ $\mu$m) shown in the bottom panel
of Fig.~\ref{fig_plot_atmosfera}, $A_{V}$ will result underestimated. As a consequence, the
$1.257$ $\mu$m line is much more suitable for deriving $A_{V}$ from low
resolution spectra. The upper panel of Fig.~\ref{fig_plot_atmosfera} indicates
that the $1.644$ $\mu$m line, as well, is bound to be slightly overcorrected most of
the times because of the absorption from lines of the 2$\nu_3$ band of CH$_{4}$
(which is centered at $1.66$ $\mu$m) and of CO$_{2}$.
In some extreme cases it may coincide with one of the two deeper lines
and underestimate the extinction. This would roughly affect the
extinction from the two line ratios in the same way.  

%%%%%%%%%%%%%%%%%%%%%%%%%%%%%%%%%%%%%%%%%%%%%
%%%%%%%%%%%%%%%%%%%%%%%%%%%%%%%%%%%%%%%%%%

Absorption from overlapping telluric lines is therefore a non-negligible source of systematic errors in
measuring NIR line ratios both from low- and high-resolution spectra. However, low-resolution emission-line 
spectra cannot be corrected to high precision in regions with a high number of atmospheric lines,
like around $1.321$ $\mu$m, whereas high-resolution spectra allow mapping the location
of telluric lines with respect to the emission lines with high accuracy through observations of
telluric standard stars or models of atmospheric transmission. In the medium- and 
high-resolution cases, 
the methods available to correct for telluric line absorption appear able to constrain the
uncertainty on the integrated line fluxes within few percent, provided the overlapping telluric
lines are faint enough to absorb $\lesssim 10$ \% of the integrated flux. More intense absorption may be difficult
to constrain below $\sim 10$ \% even in high-resolution spectra. 

\section{Deriving extinction from [FeII] line ratios}
\label{app:cal}

It can easily be found that the relationship between observed and
intrinsic line ratios, and extinction is the following:

\begin{equation}
\label{start:eq}
\log{\frac{F_{\lambda_1}}{F_{\lambda_2}}} =
\frac{A_{\lambda_2} - A_{\lambda_1}}{2.5} +
\log{\frac{I_{\lambda_1}}{I_{\lambda_2}}}
\end{equation}

where $F_{\lambda_1}, F_{\lambda_2}$ are the observed line fluxes,
$I_{\lambda_1}, I_{\lambda_2}$ are the intrinsic line fluxes, and
$A_{\lambda_1}, A_{\lambda_2}$ are the extinction values (in magnitudes)
at wavelengths $\lambda_1$ and $\lambda_2$, respectively.
If the lines originate from the same upper level, like
[FeII] $1.644$, $1.321$, and $1.257$ ${\mu}$m, then in the
optically thin case the ratio of intrinsic line fluxes is simply
given by the
ratio of the corresponding spontaneous emission coefficients, which
is constant and in principle known, multiplied by the ratio of line
central frequencies. $A_{\lambda_1}  and A_{\lambda_2}$
can be expressed in terms of $A_{V}$ by
adopting an extinction law $A_{\lambda}/A_{V}$,
where $A_{V}$ is the extinction in the photometric $V$ band.
Thus, Eq.~\ref{start:eq} can be rewritten as:

\begin{eqnarray}
\label{next:eq}
A_{V} =
\left [2.5/ \left (\frac{A_{\lambda_2} - A_{\lambda_1}}{A_{V}} \right ) \right ] \times
\left (\log{\frac{F_{\lambda_1}}{F_{\lambda_2}}} - 
\log{\frac{I_{\lambda_1}}{I_{\lambda_2}}} \right ).
\end{eqnarray}

Values of $2.5/(A_{\lambda_2} - A_{\lambda_1})/A_{V}$ are listed in Table~\ref{ext:param}
for a few different extinction laws. Table~\ref{intr:line:ra} lists a few determinations
of the intrinsic line ratios (see also the discussion in Sect.~\ref{sect:disc}). 
From Eq.~\ref{next:eq} the effect of
systematic errors $\Delta F_{\lambda}$ on the measured fluxes
is easily computed. This amounts to:

\begin{eqnarray}
\label{err:eq}
\Delta A_{V} =
\left [2.5/ \left (\frac{A_{\lambda_2} - A_{\lambda_1}}{A_{V}} \right ) \right ] \times
\nonumber \\
\left [\log \left ({1+\frac{\Delta F_{\lambda_1}}{F_{\lambda_1}}} \right ) -
\log \left ({1+\frac{\Delta F_{\lambda_2}}{F_{\lambda_2}}} \right ) \right]
\end{eqnarray}
E. g., by using the extinction law of \citet{fitzpatrick99} for $R_{V} = 3.1$
(see Table~\ref{ext:param}) and assuming $\Delta F_{\lambda_1} =0$ one obtains
$\Delta A_{V} \sim -1.2,-1.5$ for $\Delta F_{\lambda_2}/F_{\lambda_2} \sim 0.1$,
$\lambda_1 \sim 1.644$ $\mu$m and $\lambda_2 \sim 1.257, 1.321$
$\mu$m, and $\Delta A_{V} \sim 1.2, 1.5$ for $\Delta F_{\lambda_2}/F_{\lambda_2} \sim -0.1$,
$\lambda_1 \sim 1.644$ $\mu$m and $\lambda_2 \sim 1.257, 1.321$
$\mu$m. The same figures are obtained for a systematic error $\Delta X/X \sim \pm 0.1$
on the line ratio calibration factor $X$, provided that $\Delta F_{\lambda_2}/F_{\lambda_2}$
is set equal to 0 and $\Delta F_{\lambda_1}/F_{\lambda_1}$ is replaced by $\Delta X/X$. 
This shows that the accuracy of the extinction derived is extremely
sensitive to that of the measured line fluxes or line ratios.

%%%%%% TAVOLA ESTINZIONI 
\begin{table*}[!h]
\caption[]{$2.5/ (\frac{A_{\lambda_2} - A_{\lambda_1}}{A_{V}})$ for 
different extinction laws and $\lambda_1 = 1.644$ $\mu$m.}
\label{ext:param}
\begin{tabular}{lllll}
\hline
\noalign{\smallskip}
 & \multicolumn{2}{c}{$R_{V} = 3.1$} & \multicolumn{2}{c}{$R_{V} = 5.0$} \\ 
 Ext. law  & $\lambda_2 = 1.257$ $\mu$m & $\lambda_2 = 1.321$ $\mu$m & $\lambda_2 = 1.257$ $\mu$m  &
   $\lambda_2 = 1.321$ $\mu$m \\
\hline
\noalign{\smallskip}
RL85  & 23.70 & 30.46 & & \\
CCM89 & 25.48 & 32.59 & 21.99 & 28.12 \\  
F99   & 28.70 & 37.26 & 27.65 & 36.55 \\
\hline
\end{tabular}
\tablerefs{RL85: Rieke \& Lebofsky (1985);
CCM89: Cardelli, Clayton \& Mathis (1989);
F99: \citet{fitzpatrick99}; 
}
\end{table*}
%%%%%%%%%%%%%%%%%%%%%%%%%%%%%%%%%%%%%%%%%%%%%%%%%%%%%%%%%%%%%%%%%%%55

The intrinsic error $\sigma F_{\lambda}$ on the measured flux $F_{\lambda}$ propagates to give
the following error on the derived extinction:
\begin{eqnarray}
\label{int:err:eq}
\sigma A_{V} =
0.43 \times \left [2.5/ \left (\frac{A_{\lambda_2} - A_{\lambda_1}}{A_{V}} \right ) \right ] \times
\nonumber \\
\sqrt{\left (\frac{\sigma F_{\lambda_1}}{F_{\lambda_1}} \right )^{2}  +
\left (\frac{\sigma F_{\lambda_2}}{F_{\lambda_2}} \right )^{2} }
\end{eqnarray}
From Table~\ref{flux_table} and in the same assumptions as above, it can be easily 
found a range $\sigma A_{V} \sim 0.6 - 1.5$, which are comparable with the systematic errors
due to calibration. 

\section{Discussion}
\label{sect:disc}

The purpose of this section is applying the relations provided in the
previous section to our measured
Fe$^{+}$ line ratios, deriving a set of extinction values, and comparing them with
those found in the literature. This is a purely didactic exercise, since the extinction
given in the literature for our targets is a photospheric extinction, whereas the Fe$^{+}$ is likely
to arise in outermost, less reddened regions. The extinction determination to our targets is widely discussed
in \citet{hern04} (HD200775) and \citet{cohen85} (V1478~Cyg).
As for HD200775, \citet{hern04} compare optical colors and intrinsic colors
(from the star spectral type) obtaining $A_{V}$ in the range $1.8-3$ mag, which appears 
quite robust. \citet{cohen85} discuss a few diagnostics for V1478~Cyg, all pointing to an extinction 
$A_{V}$ in the range $10-10.7$ mag; values as low as $A_{V} = 8$ seems unlikely.  
Although these photospheric extinction values can only provide an upper limit to the
values yielded by [FeII] line ratios, a comparison is quite a useful exercise
for highlighting the different results from the available Einstein coefficient computations and the
effects from the various sources of errors.

We derived the extinction to the stars HD~200775 and V1478~Cyg out of the two line ratios $1.644/1.257$ and
$1.644/1.321$, from Eq.~\ref{next:eq}. 
We used the measured line ratios given in Table~\ref{flux_table} and 
the intrinsic line ratios of
Nussbaumer \& Storey (1988), as in \citet{nisini05},
the empirical determinations of \citet{giannini15}, and the values discussed
in \citet{bautista}, including their recommended values. We tested
the effect of changing the extinction law by using the relation given by 
\citet{draine89}, that of Rieke \& Lebofsky (1985) 
as in \citet{nisini05},
and that of \citet{fitzpatrick99} for $R_{V} = 5.0$.
\citet{hern04} found that most HAe/Be stars in their
sample exhibit a reddening better characterized by $R_{V} = 5.0$; the extinction curves
of \citet{fitzpatrick99} allow us to investigate the effect of changing $R_{V}$.
The results are listed in Table~\ref{t7}.
We note that only the errors due to the measured fluxes have been taken into account. Systematic
errors in the calibration coefficients (which we have estimated $\sim 5-10$ \%) 
may induce shifts of $\sim 0.5-1.5$ mag (at most).

We rule out the Einstein coefficients which lead to extinctions more than one sigma either below zero or above the
$A_{V}$ values quoted in Table~\ref{flux_table} for at least one of
the two stars. Namely, the ones labeled BPeTFDAc and Q96 \citep{bautista}.
That labeled SST can probably be discarded, as well. The upper limits of \citet{giannini15} give
a systematically lower extinction compared with the other theoretical values, therefore
we tentatively exclude them from our analysis.
The remaining sets of Einstein coefficients lead to $A_{V}$ in the range $0.3-3.6$ mag
for HD~200775, which agree with the range $1.8-3$ mag given in Table~\ref{flux_table},
and in the range $2.5-6.9$ mag
for V1478~Cyg, below the values $10-10.7$ mag given in Table~\ref{flux_table},
all systematically lower. The two line ratios always
yield reddening values consistent with each other within
errors. The differences caused by changing the extinction law adopted become important
at high reddening.

However, if we restrict our analysis to the most likely values of the Einstein coefficients,
e. g. the recommended set of 
\citet{bautista}, the lower limits of \citet{giannini15}, and the
computation by Nussbaumer \& Storey~(1988),
we find $A_{V} \sim 1-2$ mag ($R_{V} = 5$) for HD~200775, and
$A_{V} \sim 2.5-5$ for V1478~Cyg. These are systematically lower than the extinctions
given in Table~\ref{flux_table} (note that $A_{V} \sim 3$ for HD~200775
when assuming $R_{V} = 5$, as recommended by \citet{hern04}).
%It can be easily seen from Eq.~\ref{next:eq} that an error of 5 \% on the
%line-ratio correcting factors can lead to a systematic error of $\sim 0.5$ mag
%on $A_{V}$, not enough to explain the discrepancy. 
The intrinsic line ratios
entering Eq.~\ref{next:eq} are listed in Table~\ref{intr:line:ra} for the three sets
of Einstein coefficients and in the case of optically thick emission (blackbody
at 3000, 5000, 10000, 15000, and 20000 K), as well.
By comparing Eq.~\ref{next:eq} and Table~\ref{intr:line:ra} it is easily seen that 
optically thick emission would yield highly discrepant values of $A_{V}$
from the two line ratios, shifting them in opposite directions by a few magnitudes,
confirming that the line intensities are in the optically thin case for both stars. 
%If so, the Fe$^{+}$ emission may well arise from different, less extincted regions
%than the central stars. Near young stars, the [FeII] lines are usually produced
%in shocks within collimated stellar winds which usually open their ways through cavities,
%which would be consistent with less extincted [FeII] emission from the blue lobe
%of a wind from HD~200775. As for V1478~Cyg, HI and [FeII] lines are likely to
%arise in an ionized region surrounding the star, which is associated with
%an almost edge-on disk. High-resolution radio continuum maps clearly show
%an hour-glass morphology of the ionized region (White \& Becker 1985).
%The ionization potential of iron is $7.9$ eV and the critical density of the
%NIR [FeII] lines is $\sim 10^{4}$ cm$^{-3}$, so it is reasonable that the
%[FeII] lines arise in the external parts of the ionized region, which would also explain
%the lower FWHM in velocity of [FeII] lines compared to HI lines. These regions
%may well be much less extincted than the central stars.

%%%%%% TAVOLA RAPPORTI INTRINSECI 
\begin{table}[!h]
\caption[]{Intrinsic line ratios ($\frac{I_{\lambda_1}}{I_{\lambda_2}}$ in
Eq.~\ref{next:eq}) from the most recent values of the Einstein coefficients,
with $\lambda_1 = 1.644$ $\mu$m. Also listed, values in the optically thick case.
}
\label{intr:line:ra}
\begin{tabular}{lll}
\hline
\noalign{\smallskip}
 Source & $\lambda_2 = 1.257$ $\mu$m  & $\lambda_2 = 1.321$ $\mu$m \\
\hline
\noalign{\smallskip}
Rec & 0.80 & 2.91 \\
Gia15$_{min}$ & 0.83 & 3.03 \\
NS88 & 0.96 & 3.45 \\
Opt. thick A & 1.50 & 1.42 \\ 
Opt. thick B & 2.05 & 1.81 \\
Opt. thick C & 2.50 & 2.12 \\
Opt. thick D & 2.65 & 2.22 \\
Opt. thick E & 2.72 & 2.26 \\
\end{tabular}
\tablerefs{Rec: Recommended values of Bautista \\
et al.~(2015);
Gia15$_{min}$: lower limits of 
\citet{giannini15};\\
NS88: Nussbaumer \& Storey~(1988);
Opt. thick A: blackbody \\
$T= 3000$ K;
Opt. thick B: blackbody $T= 5000$ K;
Opt. thick C:\\
 blackbody $T= 10000$ K;
Opt. thick D: blackbody $T= 15000$ K;\\
Opt. thick E: blackbody $T= 20000$ K;
}
\end{table}
%%%%%%%%%%%%%%%%%%%%%%%%%%%%%%%%%%%%%%%%%%%%%%%%%%%%%%%%%%%%%%%%%%%55

%%%%%%%%%%%%%%%%%%%%%%%%%%%%%%%%%%%%%%
\begin{table*}
\caption{\label{t7} $A_V$ as a function of the adopted Einstein coefficients (first column)
for HD~200775 and V1478~Cyg. $A_V^a$ and $A_V^b$ are computed from the $1.644/1.257$ and $1.644/1.321$
line ratios, respectively.}
\footnotesize
\centering
\begin{tabular}{lcccccc}
\hline
\noalign{\smallskip}
HD~200775 & $A_{V,D}^a$  &  $A_{V,D}^b$ &  $A_{V,F}^a$ &  $A_{V,F}^b$ &  $A_{V,R}^a$  & $A_{V,R}^b$ \\
\noalign{\smallskip}
\hline
\noalign{\smallskip}
SST\tablenotemark{a}            &  $-1.0 \pm 0.5$ &  $-2.0 \pm 1.3$ & $-1.1 \pm 0.6$ & $-2.4 \pm 1.5$ & $-0.9 \pm 0.5$ & $-1.9 \pm 1.2$ \\
HFR\tablenotemark{b}            &  $0.4 \pm 0.5$ & $-0.1 \pm 1.3$ & $-0.5 \pm 0.6$ & $-0.2 \pm 1.5$ & $0.4 \pm 0.5$ & $-0.1 \pm 1.2$ \\
HFRnew\tablenotemark{c}         &  $3.0 \pm 0.5$ & $3.1 \pm 1.3$ & $3.3 \pm 0.6$ & $3.6 \pm 1.5$ & $2.8 \pm 0.5$ & $2.9 \pm 1.2$ \\
CIV3\tablenotemark{d}           &  $1.9 \pm 0.5$ & $1.9 \pm 1.3$ & $2.1 \pm 0.6$ & $2.2 \pm 1.5$ & $1.8 \pm 0.5$ & $1.8 \pm 1.2$ \\
BPeTFDAc\tablenotemark{e}       &  $-5.3 \pm 0.5$ & $-6.4 \pm 1.3$ & $-6.0 \pm 0.6$ & $-7.4 \pm 1.5$ & $-5.0 \pm 0.5$ & $-6.0 \pm 1.2$ \\
Q96\tablenotemark{f}            &  $7.9 \pm 0.5$ & $10.3 \pm 1.3$ & $8.9 \pm 0.6$ & $11.9 \pm 1.5$ & $7.4 \pm 0.5$ & $9.6 \pm 1.2$ \\
7-config\tablenotemark{g}       &  $1.1 \pm 0.5$ & $2.3 \pm 1.3$ & $1.3 \pm 0.6$ & $2.7 \pm 1.5$ & $1.0 \pm 0.5$ & $2.2 \pm 1.2$ \\
TFDAc\tablenotemark{h}          &  $0.3 \pm 0.5$ & $0.5 \pm 1.3$ & $0.4 \pm 0.6$ & $0.6 \pm 1.5$ & $0.3 \pm 0.5$ & $0.5 \pm 1.2$ \\
Rec\tablenotemark{i}            &  $1.0 \pm 0.5$ & $1.1 \pm 1.3$ & $1.1 \pm 0.6$ & $1.3 \pm 1.5$ & $0.9 \pm 0.5$ & $1.0 \pm 1.2$ \\
Gia15$_{max}$\tablenotemark{j}  &  $-0.4 \pm 0.5$ & $-0.2 \pm 1.3$ & $-0.4 \pm 0.6$ & $-0.3 \pm 1.5$ & $-0.3 \pm 0.5$ & $-0.2 \pm 1.2$ \\
Gia15$_{min}$\tablenotemark{j}  &  $0.6 \pm 0.5$ & $0.8 \pm 1.3$ & $0.7 \pm 0.6$ & $1.0 \pm 1.5$ & $0.6 \pm 0.5$ & $0.8 \pm 1.2$ \\
NS88\tablenotemark{k}           &  $2.0 \pm 0.5$ & $1.6 \pm 1.3$ & $2.2 \pm 0.6$ & $1.9 \pm 1.5$ & $1.9 \pm 0.5$ & $1.5 \pm 1.2$ \\
\hline
\noalign{\smallskip}
V1478~Cyg & $A_{V,D}^a$  &  $A_{V,D}^b$ &  $A_{V,F}^a$ &  $A_{V,F}^b$ &  $A_{V,R}^a$  & $A_{V,R}^b$ \\
\noalign{\smallskip}
\hline
\noalign{\smallskip}
SST\tablenotemark{a}            & $2.0 \pm 0.2$ & $0.8 \pm 0.8$ & $2.2 \pm 0.2$ & $0.9 \pm 0.9$ & $1.8 \pm 0.2$ & $0.7 \pm 0.7$ \\
HFR\tablenotemark{b}            & $3.4 \pm 0.2$ &  $2.7 \pm 0.8$ & $3.8 \pm 0.2$ & $3.2 \pm 0.9$ & $3.2 \pm 0.2$ &  $2.5 \pm 0.7$ \\
HFRnew\tablenotemark{c}         & $5.9 \pm 0.2$ &  $6.0 \pm 0.8$ & $6.7 \pm 0.2$ & $6.9 \pm 0.9$ & $5.5 \pm 0.2$ &  $5.6 \pm 0.7$ \\
CIV3\tablenotemark{d}           & $4.9 \pm 0.2$ &  $4.8 \pm 0.8$ & $5.5 \pm 0.2$ & $5.5 \pm 0.9$ & $4.5 \pm 0.2$ &  $4.6 \pm 0.7$ \\
BPeTFDAc\tablenotemark{e}       & $-2.4 \pm 0.2$ & $-3.6 \pm 0.8$ & $-2.7 \pm 0.2$ & $-4.1 \pm 0.9$ & $-2.2 \pm 0.2$ & $-3.3 \pm 0.7$ \\
Q96\tablenotemark{f}            & $10.9 \pm 0.2$ & $13.1 \pm 0.8$ & $12.3 \pm 0.2$ & $15.2 \pm 0.9$ & $10.2 \pm 0.2$ & $12.2 \pm 0.7$ \\
7-config\tablenotemark{g}       & $4.1 \pm 0.2$ &  $5.1 \pm 0.8$ & $4.6 \pm 0.2$ & $6.0 \pm 0.9$ & $3.8 \pm 0.2$ &  $4.8 \pm 0.7$ \\
TFDAc\tablenotemark{h}          & $3.3 \pm 0.2$ &  $3.3 \pm 0.8$ & $3.7 \pm 0.2$ & $3.9 \pm 0.9$ & $3.1 \pm 0.2$ &  $3.1 \pm 0.7$ \\
Rec\tablenotemark{i}            & $4.0 \pm 0.2$ &  $3.9 \pm 0.8$ & $4.5 \pm 0.2$ & $4.6 \pm 0.9$ & $3.7 \pm 0.2$ &  $3.7 \pm 0.7$ \\
Gia15$_{max}$\tablenotemark{j}  & $2.6 \pm 0.2$ &  $2.6 \pm 0.8$ & $2.9 \pm 0.2$ & $3.0 \pm 0.9$ & $2.4 \pm 0.2$ &  $2.4 \pm 0.7$ \\
Gia15$_{min}$\tablenotemark{j}  & $3.6 \pm 0.2$ &  $3.7 \pm 0.8$ & $4.0 \pm 0.2$ & $4.2 \pm 0.9$ & $3.3 \pm 0.2$ &  $3.4 \pm 0.7$ \\
NS88\tablenotemark{k}           & $5.0 \pm 0.2$ &  $4.4 \pm 0.8$ & $5.6 \pm 0.2$ & $5.2 \pm 0.9$ & $4.6 \pm 0.2$ &  $4.2 \pm 0.7$ \\
\end{tabular}
\tablecomments{\scriptsize $A_{V,D}$ is based on the extinction law of \citet{draine89},
$A_{V,F}$ on that of \citet{fitzpatrick99} with $R_{v} = 5.0$, and $A_{V,R}$
on that of Rieke \& Lebofsky~(1985).
The first column lists the Einstein coefficient set adopted, taken from:
(a) SST(QDZ96) \citet{bautista};
(b) HFR(QDZ96) \citet{bautista};
(c) HFR new \citet{bautista};
(d) CIV3(DH11) \citet{bautista};
(e) BP extend TFDAc \citet{bautista};
(f) Q96+4d$^2$ -corr \citet{bautista};
(g) 7-config \citet{bautista};
(h) NewTFDAc \citet{bautista};
(i) Recommended value \citet{bautista};
(j) \citet{giannini15};
(k) Nussbaumer \& Storey~(1988).}
\end{table*}
%%%%%%%%%%%%%%%%%%%%%%%%%%%%%%%%%%%%%%%%%%%%%%%%%%%%%%%%%%%%%%%%%%%%%

On the other hand, most of the Einstein coefficients yielding extinctions
inside the range limited by the photospheric value also lead to extinctions which
differ at most by $\sim 2.5$ mag. In particular, the Einstein coefficients recommended by
\citet{bautista} result in extinctions quite similar as those obtained from the lower
limits of \citet{giannini15} through observations. This points to
the latest determinations leading to uncertainties not larger than $\sim 1$ mag. 

It is interesting to investigate the discrepancies reported in the literature between
$A_{V}$ as determined from the two line ratios. 
\citet{nisini05} found that the extinction towards HH1 estimated from the $1.644/1.257$ ratio
is a factor $\sim 2$ systematically larger than that estimated from the $1.644/1.321$ ratio
when using the Einstein coefficients of Nussbaumer \& Storey (1988) 
and low spectral resolution data.
On the one hand, we note from Table~\ref{t7}
that the extinction obtained using the Einstein coefficients of  
\citet{bautista} (or the lower limit of \citet{giannini15})
is $\sim 1-2$ mag lower than that obtained using the Einstein coefficients of Nussbaumer \& Storey
(1988), which mostly removes the discrepancy pointed out by 
\citet{nisini05} on the $1.644/1.257$ ratio vs.\ optical determination. On the other hand,
the difference in extinction of the two ratios obtained using the Einstein coefficients of Nussbaumer \& Storey
(1988) is $\sim 0.5$ mag, but nonetheless within the errors propagated from
the signal to noise ratios of the single lines.

Analogously, \citet{giannini15} found intrinsic ratios
$I(1.321 \mu{\rm m})/I(1.644 \mu{\rm m})$ larger than expected from
a few literature observations of jet features (see their Fig.~3). We note that these, as well,
are mostly data from low spectral resolution spectra (SofI and ISAAC).
Low resolution spectra appear unsuitable for deriving accurate extinction
or the Einstein coefficients of the [FeII] lines due to the insufficient correction
of the atmospheric absorption achievable (see Sect.~\ref{tell:line:corr}). In this case we recommend that
the extinction be derived from the $1.644/1.257$ ratio rather than the $1.644/1.321$,
i. e. the one less affected by telluric absorption lines.

\section{Conclusions}

We have reviewed the commonly used technique of deriving extinction to young stellar objects
through [FeII] line ratios $1.644/1.257$ $\mu$m and $1.644/1.321$ $\mu$m, all lines arising
from the same upper level. In
particular, we have provided the relevant equations,
discussed all the main sources of systematic errors, and roughly tested the most recent
determinations of the Einstein coefficients available in the literature.
We have checked the method by attempting 
to derive the extinction of a Herbig Be star (HD~200775) and
a Be star (V1478~Cyg)
from high ($R\sim 50000$) resolution spectra obtained with the GIANO spectrometer. 
High resolution spectra are particularly suitable for a direct assessment of the effect of telluric absorption.
In addition, the broad band of GIANO allows simultaneous acquisition at all relevant wavelengths,
making line-ratio calibration easier. 

Although we have shown that the most recent determinations of the Einstein coefficients 
(e. g., the recommended values of \citet{bautista}) are probably accurate enough 
to derive $A_{V}$ to within 1 mag of error, we found that both the
calibration and the telluric absorption correction is critical. In addition, we have shown that
a $> 10$ \% error in the line integrated fluxes measured is enough to produce a $> 1$ mag error in $A_V$.
So, an accurate determination of reddening requires high-signal-to-noise line fluxes.
The most accurate extinction values we are able to provide for our targets are $A_V = 1.1 \pm 0.6$
for the Herbig B star HD~200775, to compare with $A_V \sim 3$ found in the literature,
and $A_V = 3.7 \pm 0.2$ for V1478~Cyg, to compare with $A_V \sim 10$ found in the literature.
This discrepancy arises from the fact that [FeII] lines originate from
less extincted regions than the central stars. 

We have also discussed the problems affecting low resolution 
($R \sim 1000- 1500$) spectra and shown that the commonly adopted
technique of correcting the telluric absorption by exploiting continuum spectra of standard stars can induce
significant calibration errors in emission line-only spectra (Herbig Haro objects, jets from young stellar objects).
In particular, the $1.321$ $\mu$m line falls in a spectral window characterized by deep absorption lines of H$_{2}$O,
so its flux is prone to be overestimated, but could also be heavily underestimated in extreme cases. If low
resolution spectra of [FeII] are used to derive $A_{V}$, we advice that this should be derived from
the $1.644/1.257$ ratio only, which is less likely to be significantly affected by the telluric absorption
correction.
Medium resolution spectra ($R \sim 10000$) appear unaffected by the problem and in any
case telluric absorption lines are easily retrieved at this resolution.  

Both the determination of $A_V$ and the empirical determination of the Einstein coefficients from NIR [FeII]
lines should be done exploiting high resolution spectra, taking particular care to calibrate and
correct the spectra for telluric absorption.

Finally, we note that the problem of the telluric correction of NIR low resolution spectra using standard stars
can affect ratios from a large
number of different lines, as well. Therefore, we advice that telluric absorption simulations (e. g., with the
ESO ASM)
be performed case by case to investigate the issue.

%% If you wish to include an acknowledgments section in your paper,
%% separate it off from the body of the text using the \acknowledgments
%% command.

%% Included in this acknowledgments section are examples of the
%% AASTeX hypertext markup commands. Use \url without the optional [HREF]
%% argument when you want to print the url directly in the text. Otherwise,
%% use either \url or \anchor, with the HREF as the first argument and the
%% text to be printed in the second.

\acknowledgments

This publication makes use of data products from the Two Micron All Sky Survey, which is a
joint project of the University of Massachusetts and the Infrared Processing and Analysis
Center/California Institute of Technology, funded by the National Aeronautics and Space
Administration and the National Science Foundation.
This work was supported by INAF through project ''Premiale E-ELT 2013''.
We also thank an anonymous referee for useful and constructive comments.

%% To help institutions obtain information on the effectiveness of their
%% telescopes, the AAS Journals has created a group of keywords for telescope
%% facilities. A common set of keywords will make these types of searches
%% significantly easier and more accurate. In addition, they will also be
%% useful in linking papers together which utilize the same telescopes
%% within the framework of the National Virtual Observatory.
%% See the AASTeX Web site at http://www.journals.uchicago.edu/AAS/AASTeX
%% for information on obtaining the facility keywords.

%% After the acknowledgments section, use the following syntax and the
%% \facility{} macro to list the keywords of facilities used in the research
%% for the paper.  Each keyword will be checked against the master list during
%% copy editing.  Individual instruments or configurations can be provided 
%% in parentheses, after the keyword, but they will not be verified.

%{\it Facilities:} \facility{Nickel}, \facility{HST (STIS)}, \facility{CXO (ASIS)}.

\appendix

\section{Data reduction}

All spectral orders obtained through a GIANO exposure are output in a single image.  To reduce  these 2D spectra,
extract the one-dimensional (1D) spectra, and wavelength-calibrate them, we used the
ECHELLE package in IRAF\footnote{IRAF is distributed by
the National Optical Astronomy Observatory,
which is operated by the Associated Universities for Research in
Astronomy, Inc., under cooperative agreement with the National Science
Foundation.} and some ad-hoc scripts gathered in the
GIANO\_TOOLS\footnote{http://www.tng.iac.es/instruments/giano/} package.
Note that a GIANO image actually contains two spectra (one per optical fiber).
Each fiber output is also split in two different light sources over the slit by using an image slicer,
so each order on the 2D subtracted image is composed of 4 close-by parallel tracks, two per fiber
(see, e. g., Fig.~1 of \citet{origlia13}).
The spectra were flat-fielded by using exposures on a continuum reference lamp and bad-pixel corrected.
A polynomial was fit to the peak positions of each (high signal-to-noise) track of the continuum-lamp
reference spectrum.
The 1D target spectra were extracted from the 2D
images by summing 4 pixels around the loci defined by the polynomials, in the direction
perpendicular to the dispersion.
For wavelength calibration we took spectra of a reference U-Ne lamp.
The resulting wavelength accuracy is better than $0.01$ nm
(see http://www.bo.astro.it/giano/documents/), i.\ e.,
better than $1.5 - 3$ km s$^{-1}$ (we note that the achieved spectral
resolution amounts to $\sim 6$ km s$^{-1}$). Finally, all four 1D spectra per order
extracted from each A$-$B image were summed together to improve the signal-to-noise ratio.
Wavelengths were converted into radial velocities in the spectral regions around the [FeII] lines
and the radial velocities were referred to the
systemic velocity so that the rest (laboratory) wavelength of each line lies
at zero velocity. This allows one to visually identify radial motions (outflows, inflows) with respect to the parent star
when plotting the spectra as a function of radial velocity. 
We searched the literature for the star radial
velocities either in the heliocentric or in the local standard of rest.
HD~200775 is a binary system with members of similar mass, so we used the relation of
\citet{alecian08} to correct velocities to the heliocentric standard.
The hydrogen emission lines appear roughly centered at 0 km s$^{-1}$ in the rest frame of
the secondary star, whereas they are slightly red-shifted in the rest frame
of the primary one. This favors the less massive member as the Herbig Be star
of the pair,
in contrast to \citet{benisty13}, who associated the Herbig Be star
with the more massive member. 
As for V1478~Cyg, radial velocities in the local standard of rest are
from \citet{gordon01}. 

\section{Notes on line flux ratio calibration}

In principle, one can correct the
1D spectra for instrumental sensitivity and continuum atmospheric absorption
by using $J, H, K$ photometry of the targets themselves from the 2MASS Point Source Catalog.
In a simple case, assuming that the source continuum spectra are well approximated by power laws, $J, H, K$
fluxes are fit by a
$F_{\lambda} = A \lambda^\alpha$ law, where $A$ and $\alpha$ are free parameters.
Then, one can use a linear least-square in logarithmic form and
calibrate the flux density for each line through
Eq.~\ref{cal:flux:line}, by measuring the continuum level (in counts) in narrow spectral
intervals adjacent to the lines. Although this entails a systematic error in
line fluxes if continuum- and line-emitting regions are not spatially coincident,
line ratios are unaffected as long as a correction for 
absorption from overlapping telluric lines is also applied where needed (see Sect.~\ref{tell:line:corr}).

However,
one has to keep in mind that Herbig Ae/Be stars are typically variable in flux. So, if their NIR spectral slopes
change as well, deriving $A, \alpha$ from the 2MASS photometry (carried out long before our observations)
is not adequate to calibrate our targets. An error in the spectral slope
enters the calibration with large effects, causing systematic errors in $A_{V}$.
By combining Eq.~\ref{next:eq} and Eq.~\ref{cal:flux:line}, and
approximating the calibration error to a systematic error $\Delta \alpha$ on the
exponent of the calibrating continuum flux, one obtains:
\begin{equation}
\label{cal:single:err}
\Delta A_{V} = \Delta \alpha \times 2.5 \log{(\frac{\lambda_1}{\lambda_2})}/
[(A_{\lambda_2} - A_{\lambda_1})/A_{V}].
\end{equation}
Using the extinction law of \citet{fitzpatrick99} for
$R_{V} = 3.1$, one obtains:
\begin{equation}
\label{red:2}
\Delta A_{V} = 3.54 \times \Delta \alpha
\end{equation}
from the $1.644/1.321$ ratio
and
\begin{equation}
\label{red:4}
\Delta A_{V} = 3.35 \times \Delta \alpha
\end{equation}
from the $1.644/1.257$ ratio. The extinction obtained
from Eq.~\ref{next:eq} is then quite sensitive to
calibration errors, as well.
%For our purpose it is also important to evaluate the error on the {\em difference}
%between the extinctions derived from the $1.644/1.257$ ($A^{a}_{V}$) and $1.644/1.321$
%($A^{b}_{V}$) line ratios, which is given by
%\begin{equation}
%\label{cal:double:err}
%\Delta A^{a}_{V} - \Delta A^{b}_{V} =  0.2 \times \Delta \alpha.
%\end{equation}
%Hence, even $\Delta \alpha \sim 1$, which roughly corresponds to a
%$K/H$ flux ratio overestimated by $\sim 40$ \%
%(or, equivalently, to $\Delta (J - H) \sim \Delta (H - K) \sim 0.3$)
%and would yield a systematic error of $\sim 3$ mag,
%would cause an extinction difference of only $\sim 0.2$ mag.

In fact, V1478~Cyg is known to be highly
variable in the optical
($\sim 2$ mag in $B$ and $\sim 1 $ mag in $R$, Gvaramadze \& Menten~(2012),
whereas HD~200775 exhibits either no significant variations in $B$ and $V$ \citep{halbedel}
or of just a few hundredth of a magnitude in the optical (Sudzius \& Sperauskas 1996).

As explained in Sect.~\ref{intercal}, direct calibration of line ratios using reference spectra
of stars of known continuum is easier and allows circumventing many problems. 
The basic relation is given by Eq.~\ref{cal:ratio:line}.
Since no observations of standard stars of known spectra were performed on the night of July 30,
a basic step is checking that the normalized 
line ratios from the star spectra taken on the nights of July 29 and 31
do not exhibit significant changes. If so, one can assume that
the continuum atmospheric absorption between telluric lines mostly depends on the air column density and that
seeing variations affect the fraction of light fed into the fibers almost independently on
wavelengths. 
We have checked the variations of $X$ (as defined in Eq.~\ref{cal:ratio:line}) in time by using
the spectra of the stars listed in Sect.~\ref{intercal}
The line-ratio calibration factors obtained from these stars are listed in Table~\ref{ratio:cal},
along with those obtained from the target continuum spectra by using their 2MASS photometry.
Table~\ref{ratio:cal} also includes calibration factors
for ratios $2.14/1.25$ $\mu$m and $2.14/1.70$ $\mu$m, wavelength intervals selected for
being devoid of telluric absorption lines. For Hip~89584, we also derived the line-ratio calibration factors
by using a blackbody to compute a scaled continuum flux. We adopted a blackbody temperature of 42280 K,
the effective temperature of an O6.5 V star according to Vacca, Garmany \& Shull (1996). The spectrum
was shifted in wavelength to account for the source radial velocity of $-53$ km s$^{-1}$. By comparing
the source NIR colors from the 2MASS Point Source Catalog and those of an unreddened O6.5 V star
as given by \citet{koor} and using the extinction law of \citet{fitzpatrick99}
for $R_{V} = 3.1$,
we obtained $A_{V} \sim 1.97$. The blackbody spectrum was then reddened by this amount, as well.
We found that the line-ratio calibration factors obtained for Hip~89584 using the reddened blackbody are within 
$1.5$ \% of those derived by simply using a fit to the 2MASS NIR fluxes. Therefore, we only
used the latter method in deriving the correction coefficients, which should be accurate enough
at least for the O stars.

%%%%% TAVOLA CON FATTORI DI CORREZIONE
\begin{table*}[!h]
\caption{Line-ratio calibration factors for selected line wavelength ratios.
Also listed, those obtained from the targets by using their 2MASS fluxes.
} 
\label{ratio:cal}
\centering
\footnotesize
\begin{tabular}{llccccc}
\hline
\noalign{\smallskip}
Target & Calibrator & Date and time & $1.644/1.257$ & $1.644/1.321$ & $2.14/1.25$ & $2.14/1.70$ \\
       &  &  of observation &   & & & \\
\hline
 & VEGA & 2013-07-29/22:14:38 &       $0.274$ & $0.340$ & $0.094$ & $0.425$ \\
 & Hip~89584 & 2013-07-29/23:27:10 &  $0.278$ & $0.340$ & $0.100$ & $0.441$ \\
 & HD~188001 & 2013-07-30/00:53:24 &  $0.269$ & $0.329$ & $0.090$ & $0.417$ \\
 & HD~190429A & 2013-07-30/01:06:38 & $0.264$ & $0.322$ & $0.096$ & $0.430$ \\
 & VEGA & 2013-07-30/03:36:33 &       $0.274$ & $0.342$ & $0.097$ & $0.422$ \\
\hline
HD~200775 & & 2013-07-31/03:41:41 & $0.394$ & $0.447$ & $0.152$ & $0.467$ \\
V1686~Cyg & & 2013-07-31/03:50:49 & $0.501$ & $0.559$ & $0.322$ & $0.712$ \\
V1478~Cyg & & 2013-07-31/04:32:31 & $0.275$ & $0.335$ & $0.100$ & $0.459$ \\
\hline
 & VEGA & 2013-07-31/20:39:41 & $0.269$ & $0.340$ & $0.086$ & $0.407$ \\
 & VEGA & 2013-08-01/03:20:47 & $0.294$ & $0.360$ & $0.124$ & $0.445$ \\
\hline
\end{tabular}
\end{table*}

%%%%%%%%%%%%%%%%%%%%%%%%%%%%%%%%%%%%%%%

Clearly, the line-ratio correction factors remained constant within at list a 10 \% during the
whole night of 29 for every wavelength pair. The picture gets worse on the night of 31.
The line-ratio correction factors to apply to the [FeII] wavelengths, however, still remained constant
within a 10 \%. Furthermore, all the night mean values are within 10 \% of each other. 
The line-ratio correction factors obtained from the 2MASS photometry of V1478~Cyg are
within 10 \% of the mean values of the other two nights, with those for the [FeII] wavelengths
within 5 \%. This suggests that its NIR spectral slope has not varied significantly. Conversely,
the line-ratio correction factors obtained from the 2MASS photometry of both HD~200775 and
V1686~Cyg are very different from all others, suggesting that their NIR spectral slopes have
significantly changed.
%These changes can be quantified by following the discussion in Sect.~\ref{app:cal}. A fit to 2MASS
%photometry measurements yields a spectral index of $\alpha = -1.1 \pm 0.05$ for HD~200775 and
%$\alpha = 1.45 \pm 0.06$
%for V1686~Cyg. By applying the line-ratio calibration factors from the night of 29
%one would obtain spectral indices $\alpha \sim -2.08$ for HD~200775 and $\alpha \sim -0.8$
%for V1686~Cyg. However, the line-ratio correction factors lead to significant residuals
%%in the spectral index of HD~200775 depending on the wavelength pair selected,
%indicating that its NIR continuum spectrum cannot be represented simply as
%an exponential function of wavelength. Such differences in spectral indices are able
%to drive significant errors in the derived extinctions, as detailed in Sect.~\ref{app:cal}. 

The small variations exhibited by the line-ratio calibration factors for different stars
throughout two nights strongly points towards confirming our assumptions of
them being only weakly sensitive to weather and seeing variations on close-by nights.
This should still allow us to obtain calibrated [FeII] line ratios with an accuracy within 5-10 \%.

\section{Telluric absorption spectra in the [FeII] region}
%%%%%%%%%%%%%%%%%%%%%%%%%%%%%%%%%%%%%%
     \begin{figure*}[h]
   \centering
   \includegraphics[width=8cm,angle=-90]{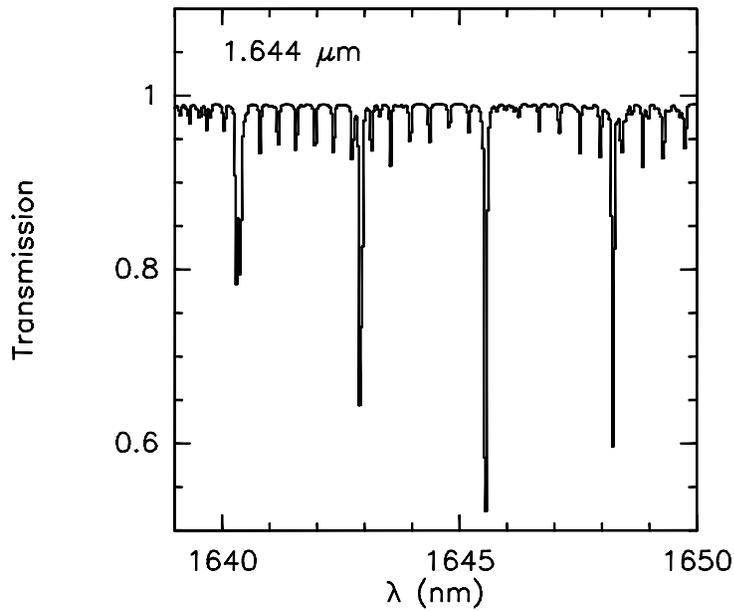}
   \caption{Atmospheric transmission around the [FeII] emission features at
$1.644$ $\mu$m computed with a spectral resolution R=50000. The absorption lines are CH$_4$
and CO$_{2}$ (the fainter ones). The spectrum is obtained with the ESO Cerro Paranal 
Advanced Sky Model.
}
              \label{atmo_wl_1}%
    \end{figure*}

     \begin{figure*}[h]
   \centering
   \includegraphics[width=8cm,angle=-90]{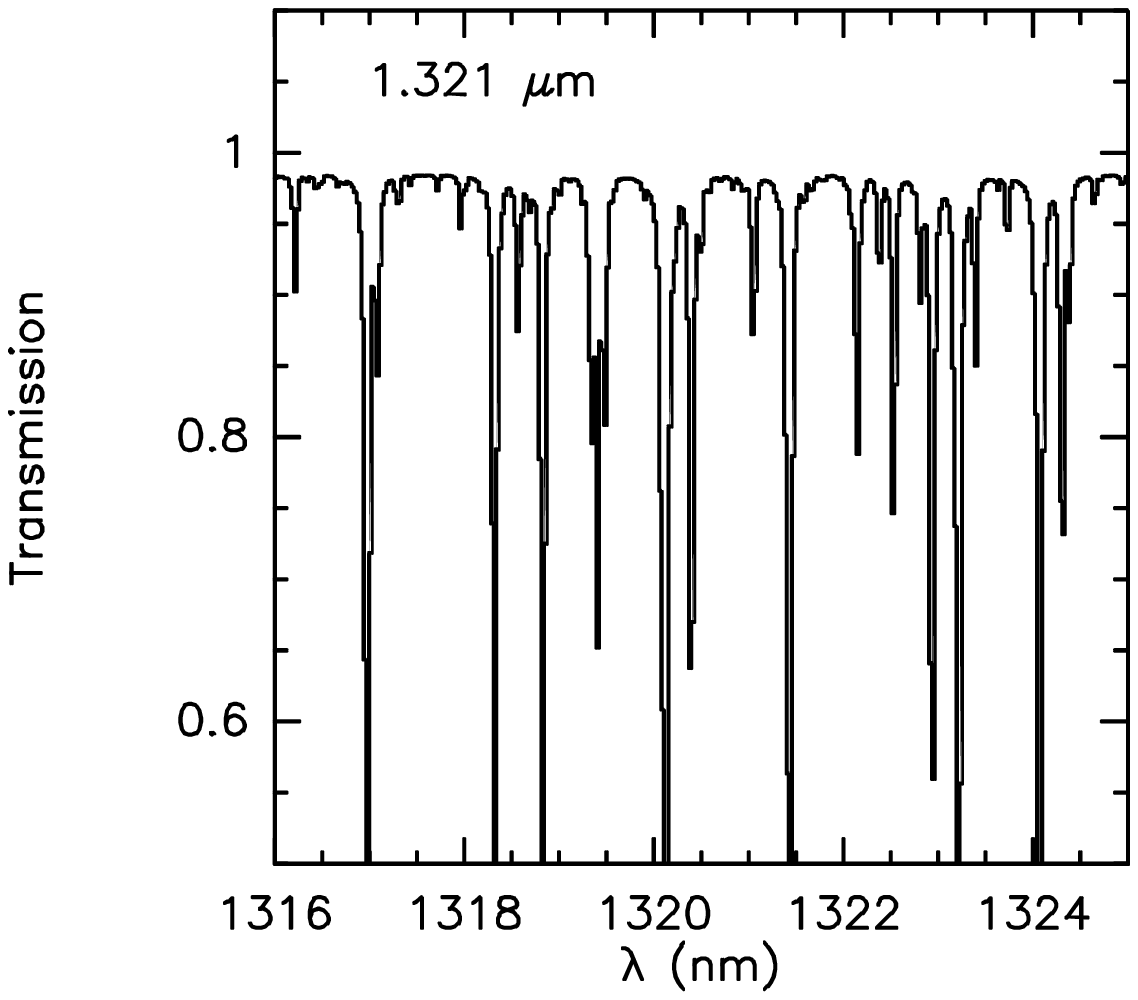}
   \caption{Atmospheric transmission around the [FeII] emission features at
$1.321$ $\mu$m computed with a spectral resolution R=50000. The absorption lines are H$_2$O. 
The spectrum is obtained with the ESO Cerro Paranal 
Advanced Sky Model.
}
              \label{atmo_wl_2}%
    \end{figure*}

     \begin{figure*}[h]
   \centering
   \includegraphics[width=8cm,angle=-90]{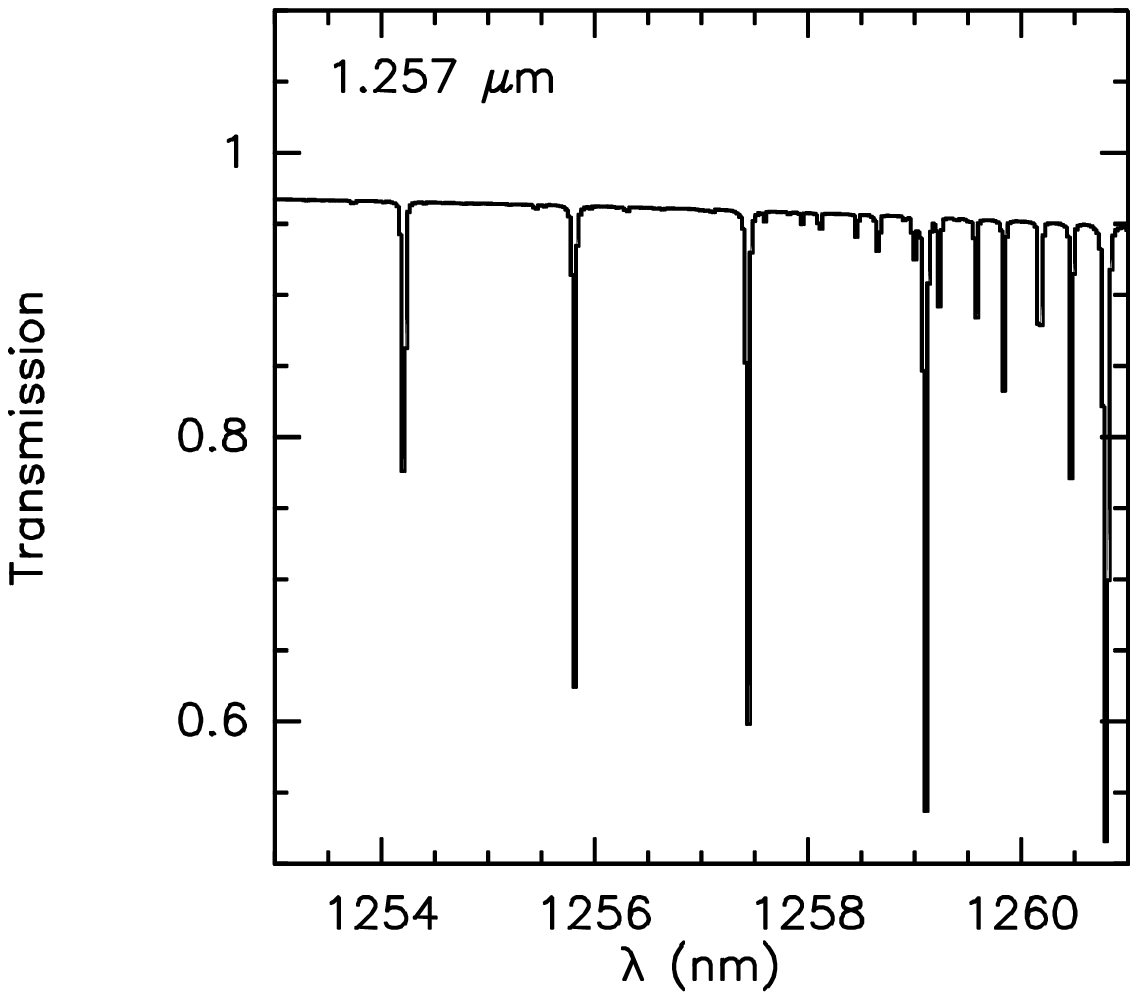}
   \caption{Atmospheric transmission around the [FeII] emission features at
$1.257$ $\mu$m computed with a spectral resolution R=50000. The absorption lines are O$_2$. 
The spectrum is obtained with the ESO Cerro Paranal
Advanced Sky Model.
}
              \label{atmo_wl_3}%
    \end{figure*}

\clearpage

\end{document}